# Exploring the Distribution for the Estimator of Rosenthal's 'Fail-Safe' Number of Unpublished Studies in Meta-analysis

**Konstantinos C. Fragkos**
*University College London, London, United Kingdom*
constantinos.frangos.09@ucl.ac.uk

**Michail Tsagris**
*Department of Computer Science*
*University of Crete, Heraklion, Greece*
mtsagris@yahoo.gr

**Christos C. Frangos**
*Department of Business Administration*
*Technological Educational Institute (T.E.I.) of Athens, Athens, Greece*
cfragos@teiath.gr

Correspondence to:
Konstantinos C. Fragkos
University College London
109 Brookhill Road
London SE18 6BJ, United Kingdom
Tel. +44 (0) 7960340489
e-mail: constantinos.frangos.09@ucl.ac.uk

# Abstract

The present paper discusses the statistical distribution for the estimator of Rosenthal's 'file drawer' number $N_R$, which is an estimator of unpublished studies in meta-analysis. We calculate the probability distribution function of $N_R$. This is achieved based on the Central Limit Theorem and the proposition that certain components of the estimator $N_R$ follow a half normal distribution, derived from the standard normal distribution. Our proposed distributions are supported by simulations and investigation of convergence.

**1.     Introduction**

Meta-analysis refers to methods focused on contrasting and combining results from different studies, in the hope of identifying patterns among study results, sources of disagreement among those results, or other interesting relationships that may come to light in the context of multiple studies (Borenstein, Hedges, Higgins and Rothstein, 2011). In its simplest form, this is normally by identification of a common measure of effect size, of which a weighted average might be the output of a meta-analysis. The weighting might be related to sample sizes within the individual studies (Whitehead and Whitehead, 1991, Hedges and Vevea, 1998). More generally there are other differences between the studies that need to be allowed for, but the general aim of a meta-analysis is to more powerfully estimate the true effect size as opposed to a less precise effect size derived in a single study under a given single set of assumptions and conditions (Rothman, Greenland and Lash, 2008).

When doing a meta-analysis, there is an increased threat of inflating publication bias (Sutton, Song, Gilbody and Abrams, 2000). Publication bias refers to the fact that statistically significant results are more likely to be submitted and published than work with null or non-significant results. This is due to pipeline bias and subjective reporting bias (Thornton and Lee, 2000). Combining published studies in meta-analysis increases the possibility that the meta-analytic output is over optimistic – and biased by publication bias (Begg and Berlin, 1988, Iyengar and Greenhouse, 1988). Methods for detecting publication include funnel plots, Begg's rank correlation test, Egger's linear regression test, Trim and Fill Method, Selection Models and Rosenthal's 'file-drawer' $N_R$ (Thornton and Lee, 2000, Kepes, Banks and Oh, 2012). Funnel plots are the most frequent method used to assess publication bias followed by Rosenthal's $N_R$ (Ferguson and Brannick, 2012).

Although Rosenthal's $N_R$ estimator of publication has been proposed as early as 1979 and is frequently cited in the literature (Rosenthal, 1979), little attention has been given to the statistical properties of this estimator. Thus, this is the aim of the present paper. The paper is organized in the following sections: initially a description of Rosenthal's $N_R$ followed by a statistical proposal for the estimator of Rosenthal's $N_R$ distribution, supported by simulations and concluding remarks.

## 2. Rosenthal's 'file drawer' $N_R$

The original and most commonly used fail-safe calculation was suggested by Rosenthal (1979), who introduced what he called the file drawer problem. His concern was that some statistically non-significant studies may be missing from an analysis (i.e., placed in a file drawer) and that these studies, if included, would nullify the observed effect. By *nullify*, he meant to reduce the effect to a level not statistically significantly different from zero. Rosenthal suggested that rather than speculate on whether the file drawer problem existed, the actual number of studies that would be required to nullify the effect could be calculated (McDaniel, Rothstein and Whetzel, 2006). This method calculates the significance of multiple studies by calculating the significance of the mean $Z$ score (the mean of the standard normal deviates of each study). Rosenthal's method calculates the number of additional studies $N_R$, with mean null result necessary to reduce the combined significance to a desired a level (usually 0.05).

The necessary prerequisites is that each study examines a directional null hypothesis such that the effect sizes $\theta_i$ from each study are examined under $\theta_i \leq 0$ or ($\theta_i \geq 0$). Then the null hypothesis of Stouffer's (1949, pp. 45) test is

$$H_0 : \theta_1 = \cdots = \theta_k = 0$$

The test statistic for this is
$$Z_S = \frac{\sum_{i=1}^{k} Z_i}{\sqrt{k}}$$

with $Z_i = \frac{\hat{\theta}_i}{s_i}$, where $s_i$ are the standard errors of $\hat{\theta}_i$. Under the null hypothesis $Z_S \sim N(0,1)$ (Rosenthal, 1978).

The number of additional studies $N_R$, with mean null result necessary to reduce the combined significance to a desired a level (usually 0.05) (Rosenthal, 1978, Rosenthal, 1979), is found after solving

$$Z_\alpha = \frac{\sum_{i=1}^{k} Z_i}{\sqrt{N_R + k}} \quad (1)$$

Hence, $N_R$ is calculated as

$$\hat{N}_R = \frac{\left(\sum_{i=1}^{k} Z_i\right)^2}{Z_\alpha^2} - k \quad (2)$$

where $k$ is the number of studies and $Z_\alpha$ is the one-tailed $Z$ score associated with the desired $\alpha$. Rosenthal further suggested that if $\hat{N}_R > 5k + 10$, the likelihood of publication bias would be minimal. Cooper (1979, 1980) called this number the failsafe sample size or failsafe *N*. If this number is relatively small, then there is cause for concern. If this number is large, one might be more confident that the effect, although possibly inflated by the exclusion of some studies, is, nevertheless, not zero (Zakzanis, 2001). This approach is limited in two important ways (McDaniel et al., 2006). First, it assumes that the effect size of the hidden studies is zero, rather than considering the possibility that some of the studies could have an effect in the reverse direction or an effect that is small but not zero. Therefore, the number of studies required to nullify the effect may be different than the failsafe *N*, either larger or smaller. Second, this approach focuses on statistical significance rather than practical or substantive significance (effect sizes). That is, it may allow one to assert that the mean effect size is not zero, but it does not provide an estimate of what the effect size might be (how it has changed in size) after the missing studies are included (Becker, 1994, Becker, 2005). Other limitations include the assumption that excluded studies show a null result whereas many may, instead, show a result in the opposite direction and the lack of a definition regarding what is a tolerable failsafe *N* value (Mullen, Muellerleile and Bryant, 2001, Aguinis, Pierce, Bosco, Dalton and Dalton, 2011).

Scargle (2000) and Schonemann and Scargle (2008) point out the fundamental critique of the fail-safe number: it treats the file drawer of unpublished studies as unbiased by assuming that their average *Z* value is zero. But if only 5% of studies that show Type I errors were published, the mean *Z* value of the remaining unpublished studies cannot be zero but must be negative. Consequently, the authors advocate that the fail-safe number is a gross overestimate of the number of unpublished studies required to bring the mean *Z* value of published studies to an insignificant level. In contrast to Rosenthal and Rubin (1978) and Rosenthal (1979), Scargle (2000, pp. 101) found out that the fail-safe number is large only if the significance level corresponding to $Z_\alpha$ of published studies is ≥2 and the publication bias probability is close to zero. Hence, the true fail safe number is almost never as large as Rosenthal's fail-safe number. Moreover, Schonemann and Scargle (2008) gave a generalization of Scargle's model, not being bound to any distributional assumptions of *Z* resulting in the same conclusions.

When comparing Stouffer's method with existing methods, it is notable that meta-analysis is usually done by identification of a common measure of effect size, of which a weighted average might be the output of a meta-analysis. The weighting might be related to sample sizes within the individual studies (Whitehead and Whitehead, 1991, Hedges and Vevea, 1998). More generally there are other differences between the studies that need to be allowed for, but the general aim of a meta-analysis is to more powerfully estimate the true effect size as opposed to a less precise effect size derived in a single study under a given single set of assumptions and conditions (Rothman et al., 2008). The two commonest methods are fixed effects and random effects models, which focus on the effect sizes from each study rather than the *Z* values – as Stouffer's method does. Testing the overall hypothesis of a significant effect is done with the Wald test on the meta-analytic mean (which is in essence a weighted mean of the effect sizes).

Stouffer's method belongs to a group of methods performing meta-analysis by pooling *p* values. These methods are summarised in Rosenthal (1978) and most commonly include Winer's method of adding the t-test statistics values (Winer, 1962), and Fisher's method of adding the logged *p-value*s (Fisher, 1948). Combining *p*-values from multiple studies has two major advantages (e.g. compared with another popular category of combining effects sizes below), including its simplicity and extensibility to different kinds of outcome variables. When the outcome variable is not binary (e.g. multi-class, continuous or censored survival), effects sizes may not be well defined, while association *p*-values can still be calculated (Tseng, Ghosh and Feingold, 2012). Limitations of these methods include vulnerability to criticisms of the individual studies being pooled, difficulty in handling the file drawer problem, and vague conclusions (Darlington and Hayes, 2000).

Hence, for many fields Rosenthal's fail-safe number remains the gold standard to assess publication bias, since its presentation is conceptually simple and eloquent. In addition, it is computationally easy to perform. In the next section we introduce necessary components to estimate its statistical distribution.

## 3. Distribution for the estimator of Rosenthal's $N_R$

First we give the definition of the folded normal distribution, needed for the estimation of Rosenthal's $N_R$ distribution.

**Definition 1:** The *folded normal distribution* is a probability distribution related to the normal distribution. Given a normally distributed random variable $X$ with mean $\xi$ and variance $\omega^2$, the random variable $Y = |X|$ has a folded normal distribution (Elandt, 1961, Leone, Nelson and Nottingham, 1961, Johnson, Kotz and Balakrishnan, 1994, Tsagris, Beneki and Hassani, 2014).

**Remark 1**: The folded normal distribution has the following properties:

a) Probability density function (PDF):

$$f_Y(y) = \frac{1}{\omega\sqrt{2\pi}} \exp\left[-\frac{(-y-\xi)^2}{2\omega^2}\right] + \frac{1}{\omega\sqrt{2\pi}} \exp\left[-\frac{(y-\xi)^2}{2\omega^2}\right], \text{ for } y \geq 0$$

b) $$E[Y] = \omega\sqrt{2/\pi}\exp(-\xi^2/2\omega^2) + \xi[1 - 2\Phi(-\xi/\omega)]$$

$$Var(Y) = \xi^2 + \omega^2 - \{\omega\sqrt{2/\pi}\exp(-\xi^2/2\omega^2) + \xi[1 - 2\Phi(-\xi/\omega)]\}^2$$

where $\Phi(\cdot)$ denotes the cumulative distribution function (CDF) of a standard normal distribution.

**Remark 2**: When $\xi = 0$, the distribution of $Y$ is a *half-normal distribution*, which is identical to the *truncated normal distribution*, with left truncation point 0 and no right truncation point. For this distribution we have

a) $f_Y(y) = \dfrac{\sqrt{2}}{\omega\sqrt{\pi}} \exp\left(-\dfrac{y^2}{2\omega^2}\right)$, for $y \geq 0$

b) $E[Y] = \omega\sqrt{2/\pi}$, $Var(Y) = \omega^2(1 - 2/\pi)$

Next, we state the following proposition.

**Proposition 1: The $Z_i$ in Rosenthal's $N_R$ estimator (expressions 1 and 2) are derived from a half normal distribution, based on a standard normal distribution.**

**Support**: When a researcher begins to perform a meta-analysis, the sample of studies is drawn from those studies that are already published. So the sample is most likely biased by some sort of selection bias, produced via a specific selection process (Mavridis, Sutton, Cipriani and Salanti, 2013). Thus, although when we study Rosenthal's $N_R$ assuming that all $Z_i$ are drawn from the normal distribution, they are in essence drawn from a truncated normal distribution. But at which point is this distribution truncated? We would like to advocate that the half normal distribution, based on a standard normal distribution is the one best representing the $Z_i$ Rosenthal uses to compute his file drawer $N_R$. The reasons for this are:

1. Firstly, to assume that all $Z_i$ are of the same sign does not impede the significance of the results from each study. That is the test is significant when either $Z_i > Z_{\alpha/2}$ or $Z_i < Z_{1-\alpha/2}$ occurs.

2. However, when a researcher begins to perform a meta-analysis of studies, many times $Z_i$ can be either positive or negative. Although this is true, when the researcher is interested in doing a meta-analysis, usually the $Z_i$ that have been published are indicative of a significant effect of the same direction (thus $Z_i$ have the same sign) or are at least indicative of such an association without being statistically significant; hence, producing $Z_i$ of the same sign but not producing significance (e.g. the confidence interval of the effect might include the null value).

3. There will certainly be studies which produce a totally opposite effect, thus producing an effect of opposite direction; but these will likely be a minority of the studies. Also there is the case that these other signed $Z_i$ are not significant.

Hence, we next discuss the statistical estimation of this distribution. We first prove the following Lemma:

**Lemma 1: The sum of k i.i.d. half normal distributions follows an asymptotic normal distribution.**

**Proof**: Let $X_i$ for $i=1,2,\ldots,i,\ldots,k$ follow a half normal distribution derived from the normal distribution $N(0,\sigma^2)$. Then, according to *Definition 1*, we have $E[X_i]=\sigma\sqrt{2/\pi}$ and $Var(X_i)=\sigma^2(1-2/\pi)$. According to the Central Limit Theorem we have:

$$\sum_{i=1}^{k} X_i \xrightarrow{d} N\left(k\sigma\sqrt{2/\pi}, k\sigma^2(1-2/\pi)\right)$$

Next, we perform the following steps to compute the distribution for the estimator of $N_R$:

**Step 1**. From *Proposition 1*, $Z_1, Z_2, \ldots, Z_i, \ldots, Z_k$ in the formula of the estimator $\hat{N}_R$ [equation (2)] are half normally distributed with $\sigma^2=1$. So according to *Lemma 1*, we have

$$S = \sum_{i=1}^{k} Z_i \xrightarrow{d} N\left(k\sqrt{2/\pi}, k(1-2/\pi)\right) \quad (3)$$

For simplicity let $\mu = k\sqrt{2/\pi}$ and $\sigma^2 = k(1-2/\pi)$. So the PDF of $S$ is

$$f_S(s) = \frac{1}{\sqrt{2\pi\sigma^2}} \exp\left[-\frac{(s-\mu)^2}{2\sigma^2}\right] \quad (4)$$

Rosenthal's expressions (1) and (2) create some ambiguity regarding the sign of $S = \sum_{i=1}^{k} Z_i$. The distribution for the estimator of Rosenthal's $N_R$ can be retrieved from a truncated version of (4) or from the folded normal distribution of (4). We describe both approaches below.

## 3.1  1st Approach: Derivation from a Truncated Normal Distribution

**Step 2a.**  From equation (1) we get that

$$S = Z_\alpha \sqrt{\hat{N}_R + k} \tag{5}$$

In this approach we advocate that Rosenthal's equations (1) and (2) implicitly impose two conditions which must be taken into account when we seek to estimate the distribution of $N_R$

$$S \geq 0 \tag{6}$$

$$\hat{N}_R \geq 0 \tag{7}$$

Expression (6) is justified by the fact that the right hand side of (5) is positive, so then $S \geq 0$. Expression (7) is justified by the fact that $N_R$ expresses the number of studies, so it must be at least $0$. Hence, expression (6) and (7) are satisfied when $S$ is a truncated normal random variable, let it be $S^*$, such that $S^* \geq Z_\alpha \sqrt{k}$. So the PDF of $S^*$ then becomes

$$f_{S^*}(s^*) = \frac{1}{\Phi(\lambda)\sqrt{2\pi\sigma^2}} \exp\left[-\frac{(s^* - \mu)^2}{2\sigma^2}\right], \; s^* \geq Z_\alpha \sqrt{k} \tag{8}$$

where $\lambda = \frac{\mu - Z_\alpha \sqrt{k}}{\sigma}$.

Then, we have

$$f_{\hat{N}_R}(n_R) = \frac{Z_\alpha}{2\Phi(\lambda)\sqrt{2\pi\sigma^2(n_R + k)}} \exp\left[-\frac{(Z_\alpha\sqrt{n_R + k} - \mu)^2}{2\sigma^2}\right], \; n_R \geq 0 \tag{9}$$

The characteristic function is

$$\psi_{\hat{N}_R}(t) = E[\exp(itN_R)] = \frac{\Phi\left(\frac{\mu_1 + \lambda}{\sigma_1}\right)}{\Phi(\lambda)} \cdot \frac{Z_\alpha \exp\left(\frac{\mu^2 it}{Z_\alpha^2 - 2\sigma^2 it} - kit\right)}{\left(Z_\alpha^2 - 2\sigma^2 it\right)^{1/2}} \quad (10)$$

where $i = \sqrt{-1}$, $\mu_1 = \frac{2\mu\sigma it}{Z_\alpha^2 - 2\sigma^2 it}$, $\sigma_1^2 = \frac{Z_\alpha^2}{Z_\alpha^2 - 2\sigma^2 it}$.

From (9) we get
$$E[\hat{N}_R] = \frac{\mu^2 + \sigma^2}{Z_\alpha^2} - k + \varepsilon \quad (11)$$

where $\varepsilon = \frac{\phi(\lambda)}{\Phi(\lambda)} \cdot \frac{\sigma(\mu + Z_a \sqrt{k})}{Z_\alpha^2}$ and $\phi(\cdot)$ is the PDF of the standard normal distribution. Also,

$$Var(\hat{N}_R) = \frac{2\sigma^2(2\mu^2 + \sigma^2)}{Z_\alpha^4} + \delta \quad (12)$$

where $\delta = \frac{\phi(\lambda)}{\Phi(\lambda)} \left[ \frac{\sigma^3(5\mu + Z_a\sqrt{k})^2}{Z_\alpha^4} - \left(\frac{\phi(\lambda)}{\Phi(\lambda)} + \lambda\right) \frac{\sigma^2(\mu + Z_a\sqrt{k})^2}{Z_\alpha^4} \right]$.

For proofs of expressions (10), (11) and (12), see Fragkos, Tsagris and Frangos (2014).

*3.2    2<sup>nd</sup> Approach: Derivation from the Folded Normal Distribution*

**Step 2b**.    In this approach we advocate that Rosenthal's equations (1) and (2) suggest that

$$S = \pm Z_\alpha \sqrt{\hat{N}_R + k} \quad (13)$$

This leads to

$$f_S(s) = \frac{1}{\sqrt{2\pi\sigma^2}} \left\{ \exp\left[-\frac{(s-\mu)^2}{2\sigma^2}\right] + \exp\left[-\frac{(s+\mu)^2}{2\sigma^2}\right] \right\}, \text{ for } s \geq 0 \quad (14)$$

Then, we have

$$f_{\hat{N}_R}(n_R) = \frac{Z_\alpha}{2\sqrt{2\pi\sigma^2(n_R + k)}} \left\{ \exp\left[-\frac{(Z_\alpha\sqrt{n_R + k} - \mu)^2}{2\sigma^2}\right] + \exp\left[-\frac{(Z_\alpha\sqrt{n_R + k} + \mu)^2}{2\sigma^2}\right] \right\}, n_R \geq -k \quad (15)$$

The characteristic function is

$$\psi_{\hat{N}_R}(t) = \frac{Z_\alpha \exp\left(\dfrac{\mu^2 it}{Z_\alpha^2 - 2\sigma^2 it} - kit\right)}{\left(Z_\alpha^2 - 2\sigma^2 it\right)^{1/2}} \tag{16}$$

and the expectation and variance are

$$E[\hat{N}_R] = \frac{\mu^2 + \sigma^2}{Z_\alpha^2} - k \tag{17}$$

$$Var(\hat{N}_R) = \frac{2\sigma^2(2\mu^2 + \sigma^2)}{Z_\alpha^4} \tag{18}$$

For proofs of expressions (16), (17) and (18), see Fragkos et al. (2014).

The exact PDF of $\hat{N}_R$ for various numbers of studies $k$ is shown in Figure 1. When the number of studies is very low, there is a discrepancy between the two densities. However, for a number of studies equal to 15 or more, the two densities have very small differences.

**Figure 1.** Exact distributions of approaches 1 and 2 for the PDFs of Rosenthal's $N_R$. The dashed line represents the PDF for approach 1 (9) and the solid line indicates the PDF for approach 2 (15).

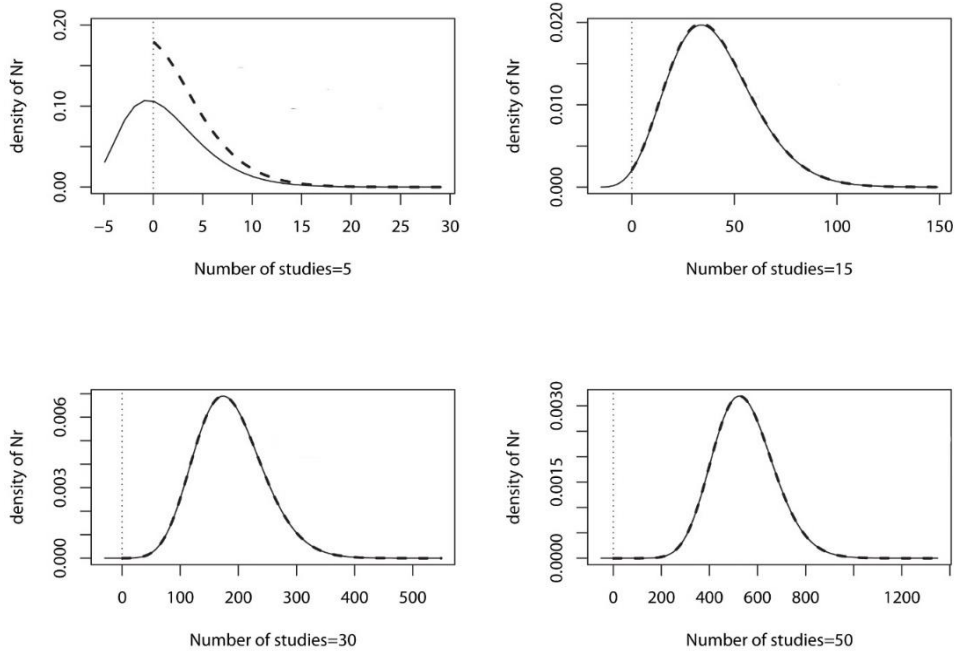

*3.3    Comments*

Each approach has advantages and limitations.

- The first approach truncates in order to satisfy Rosenthal's conditions.

- The support of $\hat{N}_R$ based on the second approach is $(-k, +\infty)$, which is contrary to the definition of $N_R$, since it allows for negative values of $N_R$.

- The PDFs of both approaches are asymptotically identical. This result extends to the expectation and variance for both approaches.

For a significantly large $n$ we have that $\Phi(\lambda) \approx 1$. So (8) becomes

$$f_{\hat{N}_R}(n_R) = \frac{Z_\alpha}{2\sqrt{2\pi\sigma^2(n_R+k)}} \exp\left[-\frac{\left(Z_\alpha\sqrt{n_R+k}-\mu\right)^2}{2\sigma^2}\right], \; n_R \geq 0 \quad (19)$$

Also from (15) we observe that $\exp\left[-\frac{\left(Z_\alpha\sqrt{n_R+k}+\mu\right)^2}{2\sigma^2}\right] \approx 0$ for a significantly large $k$. So expression (15) becomes identical to (19) with $n_R \geq -k$ in this case.

- If all the effects are insignificant and equal to zero $(Z_i = 0)$, then $N_R = -k$. So, theoretically it is possible for $N_R$ to be negative but in practice we would say it is almost surely impossible.

In the next section we present empirical results which support our results.

## 4. Simulations

In the present section we present empirical simulations computed in R (see Supplementary Materials), supporting *Lemma 1* and supporting the proposed distribution for Rosenthal's $\hat{N}_R$.

*4.1 Simulation Results for Lemma 1*

We examined the validity of *Lemma 1* empirically. For a range of sample sizes studies we generated random values from the half normal distribution. Each time we calculated the sum of these values and repeated this procedure 10,000,000 times for every chosen sample size. The histogram of these 10,000,000 sums is shown in Figure 2 along with the curve of the asymptotic normal density. We can see that the central limit theorem produces reasonable results indicating that it can be used as an approximation to the distribution of the sum of half normal variables.

**Figure 2.** Histograms of the sums of half normal variables from 10,000,000 simulations. The solid line corresponds to the asymptotic normal distribution (3).

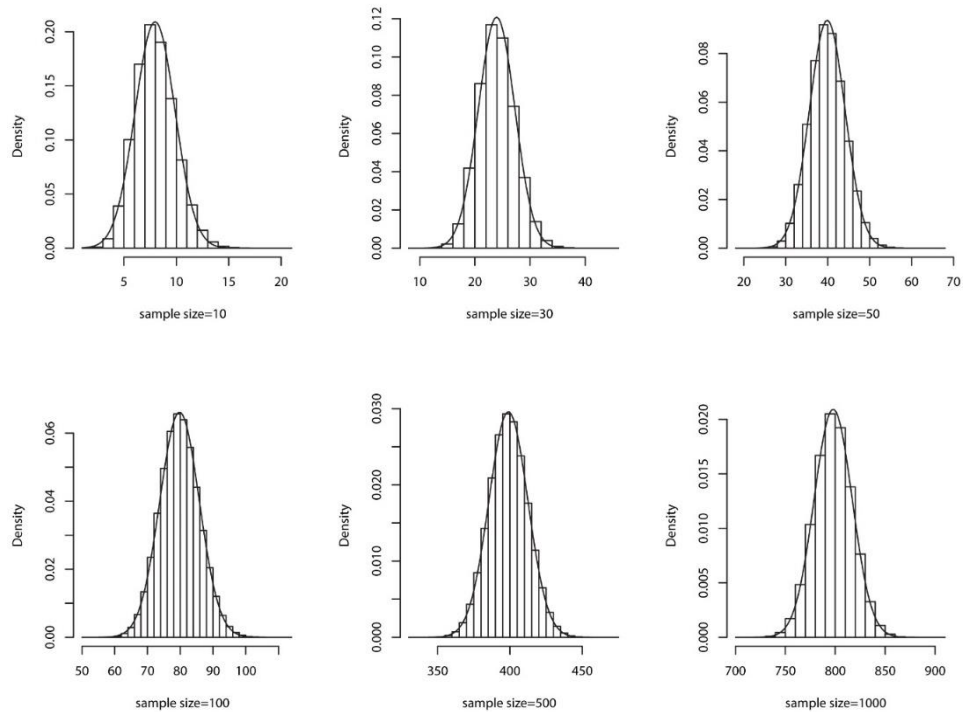

*4.2    Simulation Results for the Distribution of Rosenthal's $\hat{N}_R$*

We examined the two distributions based on simulated values of $\hat{N}_R$. For a selection of sample sizes we simulated 1,000,000 values of this estimator based on the first approach and based on the second approach. The assumption of the first approach is the condition imposed by Rosenthal; the estimator cannot take negative values. For this reason, we kept only the sums of simulated values greater than $Z_\alpha \sqrt{k}$. This leads to expression (8). The results are plotted in Figure 3. When the number of studies used in the meta-analysis is as low as 5, both distributions do not seem to provide an adequate fit for the simulated values. The first approach however seems to perform better than the second approach. Both approaches agree as the number of studies increases, starting from 15.

**Figure 3.** Histograms of 1,000,000 simulated values with $n_R \geq 0$. The dashed line represents the exact PDF for approach 1 (9) and the solid line indicates the exact PDF for approach 2 (15).

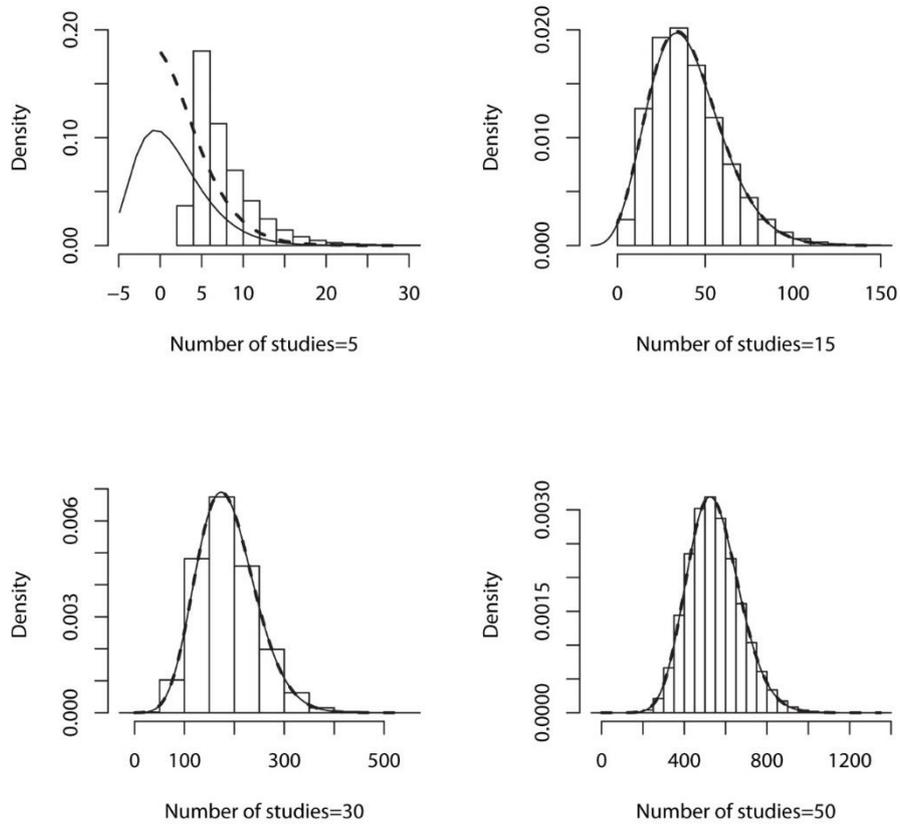

The second approach gives a range of values of $\hat{N}_R$ from $(-k, +\infty)$. The second approach does not have the non-negative values constraint, so we kept all of the simulated values. The results are shown in Figure 4, where we can see that when the number of published studies is equal to 5, the second approach fits the simulated values better. When the number of studies though is 15 or more, the two densities become indistinguishable.

**Figure 4.** Histograms of 1,000,000 simulated values with $n_R \geq -k$. The dashed line represents the exact PDF for approach 1 (9) and the solid line indicates the exact PDF for approach 2 (15).

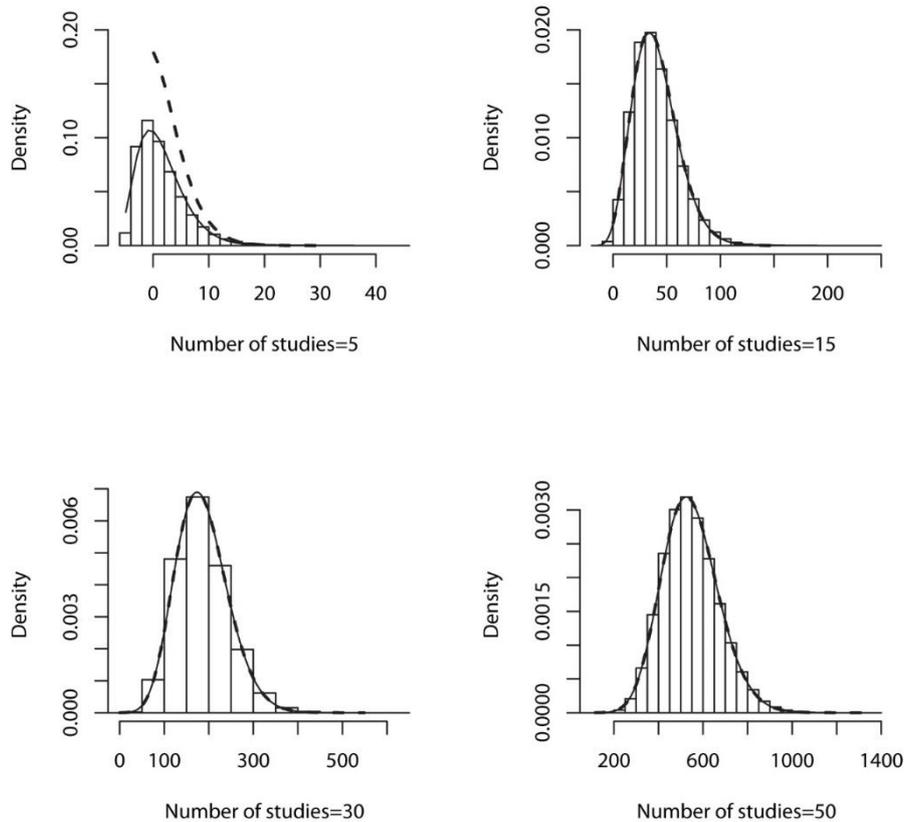

## 5. Convergence

Simulations were conducted to calculate the estimator's rate of convergence numerically. For a range of values of *k*, from 10 up to 5000 increasing by steps of 10, we performed 10,000 simulations estimating the value of $N_R$. Then, their mean value was calculated along with the absolute error

$$\left| \frac{\hat{N}_R - E[\hat{N}_R]}{E[\hat{N}_R]} \right| \tag{20}$$

where $E[\hat{N}_R]$ is the true value of the estimator as calculated using (17) and the 'hat' indicates its estimated value. It does not matter whether equation (11) or (17) is used for the true value of the estimator because the two distributions (9) and (15) become identical after 20 or 25 number of studies. The reason for using the absolute relative error and not simply the absolute error is that the estimator depends on the number of studies. The logarithm of the relative absolute error (20) against the logarithm of the number of studies is presented in Figure 5. The least squares equation is

$$\log\left|\frac{\hat{N}_R - E[\hat{N}_R]}{E[\hat{N}_R]}\right| = -4.662 - 0.539\log(k) \tag{21}$$

where $k$ is the number of studies. The slope of the decreasing line is equal to -0.539 and its 95% confidence interval is (-0.642, -0.436). Thus, we have grounds to assert that the rate of convergence of the absolute relative error is of order $O(k^{-1/2})$. Figure 5 also presents the ratio of the mean of the 10,000 estimated values from their true value $\frac{\hat{N}_R}{E[\hat{N}_R]}$ as a function of the number of studies, clearly depicting that the ratio is very close to 1.

**Figure 5.** Logarithm of the relative absolute error (20) against the logarithm of the number of studies (left) and ratio $\frac{\hat{N}_R}{E[\hat{N}_R]}$ against the number of studies (right).
.

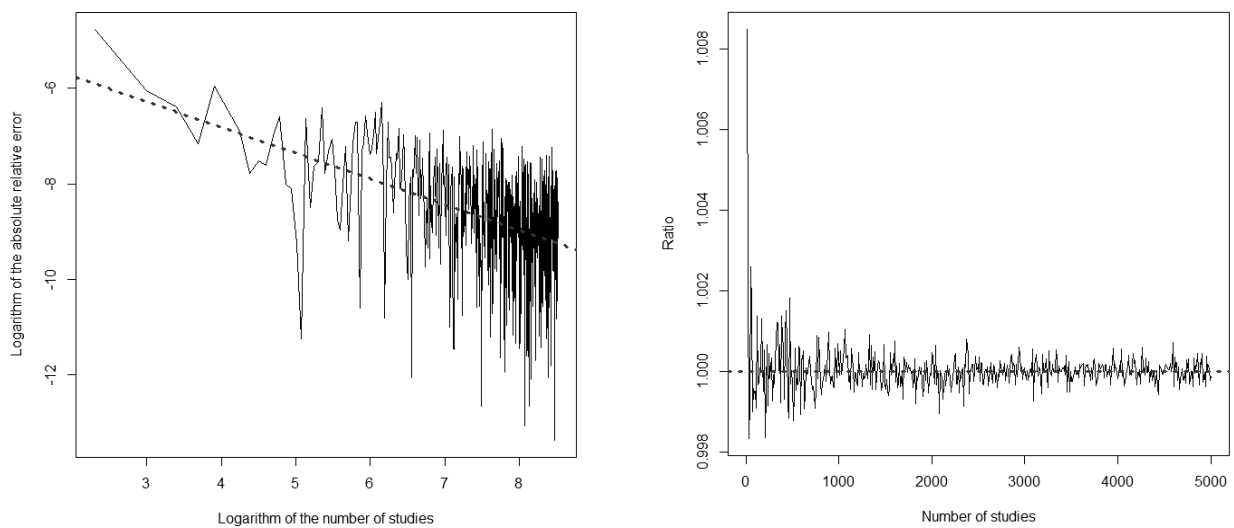

## 6. Concluding Remarks

In the present we described the statistical distribution for the estimator of Rosenthal's file drawer number. This is the first paper to provide such a description which is supported by empirical results from simulations. The estimator seems to perform efficiently when a meta-analysis includes over 15 studies, while it doesn't seem to perform optimally in meta-analyses with small number of studies analysed. The convergence of the estimator is also satisfactory and is shown (numerically) to be of order $O(k^{-1/2})$.

The present approach however has certain limitations. We based our analysis on Proposition 1, that the $Z_i$s in Rosenthal's estimator are half normally distributed. First of all, this might inflate the estimator, since it considers only the positive values of $Z_i$. Secondly this might not be most optimal distribution to sample from. We are currently performing research considering that the $Z_i$s are retrieved from a skew normal distribution (Azzalini and Dalla Valle, 1996, Azzalini and Capitanio, 1999). This approach might require fewer assumptions and the dynamic properties of the skew normal distribution might allow for more general results.

**Acknowledgements**

We thank Ioannis Pantazis for constructive feedback during the writing of the manuscript. We would also like to thank the anonymous reviewers for the valuable comments and suggestions which helped to improve the quality of this paper.